# Ab-initio approach to the stability and the structural, electronic and magnetic properties of the (001) Znfe$_2$O$_4$ surface terminations.

K. L. Salcedo Rodríguez[1], J. J. Melo Quintero[1], H. H. Medina[1], A. V. Gil Rebaza[1,2], C. E. Rodríguez Torres[1] , and L. A. Errico [1,3]

[1]IFLP y Departamento Física, Facultad de Ciencias Exactas, Universidad Nacional de La Plata-CCT La Plata CONICET, C.C. 67, CP 1900, La Plata, Argentina.

[2] Grupo de Estudio de Materiales y Dispositivos Electrónicos (GEMyDE), Dpto. de Electrotecnia, Fac. de Ingeniería, UNL.

[3] Universidad Nacional del Noroeste de la Provincia de Buenos Aires – UNNOBA, Monteagudo 2772, Pergamino, CP 2700, Buenos Aires, Argentina.

R. Faccio[4]

[4]Centro NanoMat-DETEMA, Facultad de Química, Universidad de la República, Uruguay.

W. A. Adeagbo[5] and W. Hergert[5]

[5]Institut of Physik, Martin-Luther-Universitaet Halle-Wittenberg, Von-Seckendorff-Platz 1, 06120 Halle, Germany

**Abstract**

We present a Density Functional Theory (DFT) based study of the structural and magnetic properties of the (001) surface of the semiconducting oxide ZnFe$_2$O$_4$ (spinel structure). The calculations were performed using the DFT based *ab initio* plane wave and pseudopotential method as implemented in the Quantum Espresso code. The all electron Full-potential linearized-augmented-plane-wave method (FP-LAPW) was also employed to check the accuracy of plane wave method. In both calculations the DFT+*U* methodology was employed and different (001) surface terminations of ZnFe$_2$O$_4$ were studied: We find that the surface terminated in Zn is the stable one. For all the (001) surface terminations our calculations predict that the Zn-Fe cationic inversion (antisites), which are defects in bulk ZnFe$_2$O$_4$, becomes stable and an integral part of the surface. Also, a ferrimagnetic behavior is predicted for the case of antisites in the superficial layer. Our results for different properties of the surface of ZnFe$_2$O$_4$ are compared with those obtained in bulk samples and those reported in the literature.

# 1. Introduction

Due to their electronic, optical and magnetic properties ferrites ($XFe_2O_4$ X: Zn, Co, Ni, Al, Mg, Ti, [1]) are very interesting materials both from a fundamental and application point of view. In the last decades, progress in synthesis process renews the interest in this kind of insulating oxides and improves their physical properties, expanding their applications to new areas, for example, spintronics [2-4]. In particular, Zinc ferrite ($ZnFe_2O_4$) is of interest not only in basic research due to its intriguing magnetic behaviour [5-11] but also has great potential in technological applications [12-15].

Ferrites crystallize in the spinel-type structure, in which the cations occupy tetrahedral and octahedral sites, the so-called A and B sites [1]. Two types of ferrites can be distinguished: the normal ferrites and the partially inverted ones. In the first case, the X ions occupy the A sites and the Fe atoms the B ones. In the second case, some percentage of the X atoms is located at the B sites instead of the A sites and corresponding concentration of Fe then occupies A sites. This cation distribution is described with the formula $(X_{1-\alpha}Fe_\alpha)[Zn_\alpha Fe_{2-\alpha}]O_4$ where round and square brackets denote A and B sites, respectively and $\alpha$ is the inversion parameter. For $\alpha = 1$, the inverted structure is obtained. $ZnFe_2O_4$ belongs to the class of normal ferrites, but, depending on the sample preparation method and on the thermal treatment, some degree of inversion could appear. This is important, because the magnetic properties of ferrites strongly depend on the cation distribution in the A and B sites. Magnetic coupling in ferrites occurs via super-exchange between the X and Fe cations via intermediate oxygen ions, giving rise A–O–A, B–O–B and A–O–B couplings in which A-O-B interaction is much stronger than the others. In no-defective bulk $ZnFe_2O_4$ the absence of Fe atoms at A sites (populated by non-magnetic Zn atoms) results in weak antiferromagnetic exchange interactions, making ZFO a strongly paramagnetic oxide, with antiferromagnetic coupling only below about 10K [16] and a quite complex ground state [17, 18 and Refs. therein]. However, when particle size is reduced to the nano-scale some degree of inversion that may scale with the surface area of the samples is observed [19-21]. The presence of inversion gives rise to A–O–B coupling, enhancing the magnetic interactions and the ferrimagnetic behavior at high temperatures [20].

In this work, our aim is to understand, by means of *ab initio* calculations, the structure at the atomic scale (which has remained unsolved) and the stable termination of the (001) surface of $ZnFe_2O_4$. Additionally, it is relevant to know how the surface termination affects the structural and electronic properties of the near surface layers and the cation inversion process.

**1. Calculations Details**

$ZnFe_2O_4$ crystallizes in the spinel structure, space group $Fd3m$ ($Oh7$), with Zn and Fe ions at two different crystallographic sites (sites A and B). These A and B sites have O4 (tetrahedral) and O6 (octahedral) oxygen coordination, respectively [1] and the unit cell contains eight formula units. The spinel structure is characterized by two parameters, the lattice constant $a$ and the oxygen position parameter $u$. Different experimental determinations of the lattice parameter $a$ have been reported so far, ranging from 8.43 to 8.46 Å [8, 22-25]. $ZnFe_2O_4$ adopts the normal spinel structure, in which the Zn atoms are located at tetrahedral A-sites (Wyckoff position $8a$), whereas the Fe atoms (which carry a magnetic moment due to the partially filled $3d$ shell) occupy the octahedral B-sites only ($16d$). The O atoms are at $32e$ Wyckoff positions with $u$=0.258, Ref. 25. Our aim here is to study the (001) surface of $ZnFe_2O_4$ and how the surface affects the formation of defects (cationic inversion and oxygen vacancies in the present case). Two types of (001) surface terminations can be obtained from a bulk truncation of $ZnFe_2O_4$: the fully Zn-terminated surface (Zn2) or an O4-Fe4-O4 termination that "exposes" O and Fe atoms. A third termination was studied, a reduced O4-Fe4 termination. These surface terminations are shown in figure 1. In order to study these (001) surface terminations *ab initio* calculations have been performed on the basis of Density Functional Theory (DFT) using two different methods of calculation, the Full-Potential Linearized Plan Waves method (FP-LAPW) [26-28] as implemented in Wien2k code [29] and the plane wave and pseudopotential method as implemented in Quantum-Espresso code (QE, Refs. 30 and 31). The first method is recognized as one of the more accurate and precise for the determination of the electronic structure of solids but is computationally expensive and time consuming when large systems or systems that present low symmetries or reduced dimensions (as in the case of surfaces). The plane wave and pseudo-potential methods are more flexible to study structural and electronic properties of complex systems, but their computational accuracy must be study.

In the case of the *ab initio* plane wave and pseudopotential calculations the exchange and correlation effects were treated within density-functional theory using Perdew-Burke-Ernzerhof (PBE) parameterization of the generalized gradient approximation (GGA) [32]. Since the exchange and correlation effects included in GGA are insufficient to describe 3-$d$ transition oxides, GGA plus the Hubbard $U$ term (GGA+$U$) was employed [33]. In this study we took $U = 5$ eV for the Fe-3$d$ orbitals. This value was selected after the study of a set of iron oxides. The ionic cores were described using ultrasoft pseudo-potentials from the Standard Solid State Pseudopotentials library (SSSP, Ref. 34), where the converged kinetic energy cutoff for the wavefunction and charge density were set to 70 Ry and 500 Ry, respectively. For the electronic integration, the irreducible Brillouin zone was described according to the Monkhorst-Pack scheme [35] using a 2x2x1 $k$-point grid. Atomic positions were optimized until obtain forces acting on all ion were below 0.05 eV/Å. This tolerance value was taken because atomic displacements produced by forces smaller than 0.05 eV/Å results in changes in energies differences, magnetic moments or bond-lengths that are below our convergence error.

To validate the accuracy of the QE calculations, scalar relativistic FP-LAPW calculations were performed. Exchange and correlation effects were also treated within density-functional theory using the GGA+$U$ formalism with $U = 5$ eV for the Fe-3$d$ orbitals [17, 36]. The atomic sphere's radii used for Zn and Fe were 1.06 Å and for the oxygen atoms the radii were 0.8 Å. The parameter $R_{MT} K_{max}$ controls the size of the basis set and was set to 7 ($R_{MT}$ is the smallest muffin-tin radius and $K_{max}$ the largest wave number of the basis set). We introduced local orbitals to better describe O-2$s$, Fe-3$p$ and Zn-3$p$ orbitals [29]. Integration in reciprocal space was performed using the tetrahedron method, taking a 2x2x1 $k$-point mesh to map the first Brillouin zone. Structural distortions were computed following the procedure reported in [37]. Atomic position has been optimized until the forces over each ion were below the tolerance criterion of 0.05 eV/Å.

We modeled the (001) surface terminations by using the slab-supercell approach. For the simulation of each termination we considered three different periodic slabs. Initially we study a slab composed by nine layers of Zn2 and O4-Fe4-O4 stacked alternately, separated by a vacuum space, see Figure 1. Then, the unit cells used to simulate the fully terminated Zn2, O4-Fe4 and O4-Fe4-O4 (001) terminations contains 58, 60 and 68 atoms, respectively. The thickness of these slabs is 8.5 Å. Our calculations showed that 11 Å of vacuum is sufficient to decouple the slabs. FP-LAPW and QE calculations were performed for this 8.5 Å thickness slab. To ensure that artificial interaction between the upper and lower layers of the slab is negligible, QE calculations considering a 17.0 Å thickness slab were also performed. This slab is composed by 17 layers of Zn2 and O4-Fe4-O4 stacked alternately (Figure 2). The slabs contain now 114, 116 and 124 atoms in the cases of the Zn2, O4-Fe and O4-Fe4-O4 (001) surface terminations. These larger slabs were studied performing QE calculations only.

## 2. Results and Discussion

Initially, we briefly report the results obtained for bulk $ZnFe_2O_4$. The lowest energy configuration corresponds to an antiferromagnetic system that consists of a pair of spins ferromagnetically aligned but antiferromagnetically to another pair from the nearest layer [17]. For this arrangement our calculations predict $a$=8.46 Å, $u$=0.2600 and magnetic moments at the Fe sites, $\mu$(Fe), of ±4.20 $\mu_B$ (no spin polarization was found at the Zn sites, while the magnetic moments at the oxygen atoms are smaller than ±0.05 $\mu_B$). These theoretical results are in excellent agreement with the experimental ones [see Ref. 17]. This spin configuration will be used as the starting one for our study of the three (001) surface terminations described in the previous section.

We also study the role of defects in bulk $ZnFe_2O_4$. We have considered the Zn-Fe anti-site defect formed by cation inversion between Zn and Fe. In this case, FP-LAPW calculations were performed swapping one Zn atom with one of its first neighbor Fe atom. FP-LAPW predicts that the magnetic moment of the Fe atom at the A-site is in the order of +4.2 $\mu_B$, the same value that we obtained for Fe at the B-sites in the pristine structure. The Fe atoms that remain at the B-site present magnetic moments of ±4.2 $\mu_B$. O atoms, the Zn atoms at A-site and the B-site are not spin-polarized and the system is still antiferromagnetic. But this is not the lowest energy magnetic configuration of partially inverted $ZnFe_2O_4$. The lowest energy configuration corresponds to the case in which a Fe atom first neighbor to the Fe at the A-site changes its spin orientation, giving rise to a cell with a net magnetic moment of 0.6 $\mu_B$/Fe atom (ferromagnetic system). Finally, the normal structure has the lowest energy, in agreement with the literature [Ref. 36 and references therein]. In the case of bulk $ZnFe_2O_4$, we obtained $\Delta E=E^{normal}-E^{inv}$=-85 meV per unit formula (*u.f.*), indicating that the first structure is energetically favorable.

The presence of an oxygen vacancy (reduced $ZnFe_2O_4$) produces changes in the magnetic moments. The three Fe atoms nearest neighbors to the vacancy sites present magnetic moments of +4.0 $\mu_B$ (two Fe atoms) and -3.6 $\mu_B$ for the remaining Fe atom. All the other Fe atoms of the cell present magnetic moments of ±4.2 $\mu_B$, as in the case of the stoichiometric structure. Additionally, spin polarization c.a. ~0.1 $\mu_B$ at the O sites is predicted (the Zn atoms remains non-polarized), see Ref. 36. Reduced $ZnFe_2O_4$ with normal structure has lower energy than the structure with partial inversion. However, in the case of reduced $ZnFe_2O_4$, the energy difference between the normal and partially inverted structures is -36 meV/*u.f*. This result shows that the presence of oxygen vacancies (reduced coordination) in $ZnFe_2O_4$ could favor the cationic inversion. When an oxygen vacancy and one antisite in the bulk structure is considered, the lower energy structure corresponds to an oxygen vacancy first neighbor to Zn in site B, but not first neighbor to Fe in site A (oxygen vacancy at 3.5 Å of Fe at site A, see Ref. 36). For this reduced and partially inverted structure the lowest energy magnetic configuration corresponds to a ferrimagnetic system with a net magnetic moment of 1.25 $\mu_B$/Fe atom, showing that defects (antisites and reduced oxygen coordination) play a fundamental role in the formation of local ferromagnetic coupling between Fe ions, giving rise to a ferrimagnetic ordering in this otherwise antiferromagnetic oxide,

Now we can discuss our results for the (001) surface terminations of $ZnFe_2O_4$. We will start our discussion with the QE and FP-LAPW results obtained for the cases of the 8.5 Å thickness slab. QE and FP-LAPW predict the same results for the surface structure: in the cases of the Zn2 and O4-Fe2-O4 a strong surface reconstruction occurs. For the Zn2 surface termination the superficial Zn atoms relax inward the sub-surface layer, being the magnitude of the Zn atoms displacement of about 1 Å (see Figure 1.d.). In the case of the O4-Fe4-O4 termination, the Zn atoms located in the sub-surface layer (1 Å below the O4-Fe4-O4 superficial layer) relax outwards to the surface. The magnitude of the Zn atoms displacement along the *z*-direction is c.a. ~0.5 Å (see Figure1.e). As a conclusion, we can point out that our *ab initio* calculations predict that, after the surface reconstruction, both terminations yield a similar structure with a superficial layer O4-Fe4-Zn2-O4. In the case of the reduced O4-Fe4 surface termination, the surface reconstruction is less pronounced (see Figure 1.f.). QE calculations predict very similar results for the structural properties of the (001) surface terminations of ZFO when a slab of 17 Å thickness is considered.

In order to determine the stability of each surface termination we study the surface energy formation that was calculated as [38, 39]:

$$\gamma = \frac{E_{slab} - [N_{Zn}\mu(\text{Zn}) + N_{Fe}\mu(\text{Fe}) + N_O\mu(\text{O})]}{2A}$$

(1)

to take into account the different stoichiometries of each termination. In the above equation $E_{slab}$ is the total energy of the slab structure considered $N_{Zn}$, $N_{Fe}$, and $N_O$ are the number of Zn, Fe and O atoms in the slab (the number of Zn. Fe, and O atoms for each slab configuration is shown in Table 1). $\mu(\text{Zn})$, $\mu(\text{Fe})$ and $\mu(\text{O})$ are the chemical potentials of Zn, Fe, and O that must satisfy:

$$E(\text{ZFO}) = \mu(\text{Zn}) + 2\mu(\text{Fe}) + 4\mu(\text{O})$$

(2)

and

$$E(\text{ZnO}) = \mu(\text{Zn}) + \mu(\text{O})$$

(3)

being $E(\text{ZFO})$ and $E(\text{ZnO})$ the total energies of one unit formula of bulk $ZnFe_2O_4$ and ZnO, respectively.

The chemical potentials vary depending on the experimental environments in which the surface is growth, but there are upper and lower limits that define a range for the chemical potentials. The upper limit of $\mu(\text{O})$ is the chemical potential of an oxygen molecule, $\mu(O_2)$. For $\mu(\text{O})$ larger than $\mu(O_2)/2$ vaporization of oxygen as molecules from the $ZnFe_2O_4$ surfaces will occur. In the following we call this upper limit "O-rich environment". In this situation, $\mu(O^{rich})=\mu(O_2)/2$. The lower limit of $\mu(\text{O})$ correspond to the upper limits of $\mu(\text{Zn})$ and $\mu(\text{Fe})$ which is the Zn and Fe chemical potentials in metallic Zn and Fe. For larger values of $\mu(\text{Zn})$ and $\mu(\text{Fe})$ segregation of metallic Zn and Fe will occur. This is the "oxygen-poor environment".

In order to obtain the required chemical potentials FP-LAPW and QE calculations (with the same precision than those corresponding to the slab structures and bulk $ZnFe_2O_4$) for a $O_2$ molecule and metallic Zn, metallic Fe and ZnO were performed. It was found that $\mu(O^{poor})-\mu(O^{rich})=-5.4$ eV for the present calculations. By using eqs. 1-3 and the surface energy for the O-rich environment ($E^{O\ rich}$) we can write the surface energy as a linear function of the oxygen chemical potential:

$$\gamma = \frac{E^{O\,rich}}{2A} + \frac{\left[\frac{3}{2}N_{Fe} + N_{Zn} - N_O\right]}{2A}\Delta\mu$$

(4)

With $\Delta\mu=(\mu(O)-\mu(O^{rich}))$. From detailed convergence studies we determine that our precision error in $\gamma$ is in the order of 5 meV/Å$^2$.

In a first stage we will analyse the stability of the normal (001) terminations. Figure 3 shows the calculated surface energies $\gamma$ as a function of the oxygen chemical potential for the case of the FP-LAPW calculations (the QE calculations follow the same trend). It is predicted that the Zn2 surface termination has the lowest surface energy formation for the possible range of the oxygen chemical potential. In Table 2, the values of $\gamma$ for the extreme oxygen potential range (O-rich and O-poor) are shown. When the 17 Å thickness slabs are considered as a model for the simulation of the (001) terminations of ZnFe$_2$O$_4$ QE calculations predicts a similar trend, i.e, Zn2 is the stable termination. Calculations performed using as a model for the surface a 25.5 Å thickness slab and 16 Å of vacuum confirm all the discussed results for the surface stability and confirms that even for the 8.5 Å slab thickness the spurious interaction between the upper and lower layers of the slab is negligible. As can be seen in Figure 3, the O4-Fe4 termination is more stable than the O4-Fe4-O4 in the O-poor environment. The inverse situation is observed, as expected, for the O-rich condition. The formation of additional oxygen vacancies is energetically unfavorable even in the O-poor environment.

In Figure 4 we present the QE results for the magnetic moments at the Fe sites for the case of the 17 Å thickness slab as a function of the distance of the Fe atom to the surface for the three terminations studied. As a general result we can see that Fe atoms in the surface present magnetic moments smaller than those located in the bulk. For the case of the O4-Fe4 and the O4-Fe4-O4 terminations, the magnetic moment of the Fe atoms located in the superficial layer are 3.95 and 3.60 $\mu_B$, respectively. This reduction of the magnetic moments of the Fe atom located in the superficial layers of both terminations is associated to the change in the O-coordination of these Fe atoms and the subsequent changes in the charge in the atomic spheres of these Fe atoms (the reduction of the oxygen coordination of the Fe atoms at the surface induces an increment in the charge in this Fe atoms from 23.7$e$ up to 23.9$e$). A similar effect was found in reduced bulk ZnFe$_2$O$_4$ for the first neighbour's Fe-atoms of an oxygen vacancy site [36]. The Fe atoms located in layers 2.3 Å (at least) below the surface present magnetic moments in the order of 4.10-4.20 $\mu_B$, the same value obtained for bulk $ZnFe_2O_4$. For both O4-Fe4 and O4-Fe4-O4 terminations no spin-polarization was found at the Zn-sites while spin-polarization in the order of 0.15 $\mu_B$ was found for some oxygen atoms (a similar result was obtained in reduced $ZnFe_2O_4$).

In the case of the Zn2 surface termination, a different behaviour was found. Similar to the previous cases, the Fe atoms located 1.3 Å below the surface present a spin-polarization of 4.00 $\mu_B$, an expected reduction (due to the smaller oxygen coordination of these Fe atoms) compared to the case of bulk $ZnFe_2O_4$. But, the Fe atoms located 3.4 Å below the surface present even smaller magnetic moments, 3.60 $\mu_B$. This result can be explained in terms of the superficial reconstruction and the charge rearrangement that induces: when the unreconstructed Zn2 surface is inspected magnetic moments of ±4.2 $\mu_B$ is predicted for Fe-atoms located at 3.4 Å from the surface. After the structural reconstruction (that implies a large displacement of the superficial Zn2 atoms to the O4-Fe4-O4 sub-surface layer) the magnetic moment of the Fe atoms of the sub-surface layers drops to ±3.60 $\mu_B$. For this Zn2 termination, no magnetic moments were found at the Zn sites and some oxygen atoms present magnetic moments in the order of 0.10 $\mu_B$.

It is important to mention here that QE and FP-LAPW calculations performed in the 8.5 Å thickness slab predicts the same behaviours previously discussed for the magnetic moments.

After the study of the slab with a normal distribution of Zn and Fe cations in the A and B sites of the spinel structure of $ZnFe_2O_4$, we have considered the Zn-Fe antisite defect formed by cation inversion between superficial and deep Zn and Fe atoms. Our DFT-based calculations show that, independently of the surface termination considered, the normal structure and those with a superficial antisite have the same energy (differences smaller than the convergence error), see Table 2. This result is contrary to those found in bulk $ZnFe_2O_4$, for which the inversion process is strongly endothermic. The superficial cation inversion is promoted by the rearrangement of the electronic structure in the surface layer in order to compensate the changes originated by the surface. As we discussed previously, a similar effect is observed in volumetric and reduced $ZnFe_2O_4$: the energy necessary to produce a Zn-Fe antisite is 60% smaller than those required in the case of stoichiometric $ZnFe_2O_4$ [36]. Concerning magnetic moments at the Fe sites, the same results previously discussed for the normal structures were found. The only difference is that the magnetic moments of the Fe atoms allocated at the A-sites are 0.1-0.2 $\mu_B$ smaller than those of the Fe atoms that remain at the B-sites. This diminution is observed for Fe atoms allocated at the A sites in the surface layer or even in deep layers.

Based in our results we can conclude that Zn-Fe inversion is not a trapped defect (as in the bulk case) but also an integral component of the surface of $ZnFe_2O_4$. This conclusion is in agreement with experimental results that shown that the degree of cation inversion may scale with the surface area in $MgAl_2O_4$, $ZnFe_2O_4$, and $NiFe_2O_4$ ferrites spinel powders [21] or thin films [40]. In other works, it was shown that $ZnFe_2O_4$ nanoparticles of 6 nm diameter present a high degree of inversion that is associated to surface effects, i.e., involves cations allocated at the surface layer [41, 42]. A similar result was obtained by Rasmussen *et al*. [37] in $MgAlO_4$ spinel, suggesting that surface-induced inversion could a general phenomenon in spinel-type oxides.

To conclude, the study the magnetic configuration of the Zn2 surface. First results show that a ferrimagnetic structure (with a net magnetic moment of 1.2 $\mu_B$/Fe atom) is the lowest energy configuration. This result indicates that the surface (reduced Fe-coordination) favours the cationic inversion and in turns the cationic inversion favours ferrimagnetic configurations. A systematic study of the magnetic behaviour of the surface terminations of ZnFe2O4 is now in progress.

## 3. Conclusions.

In this work we have study, by using two different DFT-based methods (all electron FP-LAPW and plane wave and pseudopotentials method as implemented in the Quantum Espresso code) the stability of the Zn2, O4-Fe4 and O4-Fe4-O4 terminations of the (001) surface of $ZnFe_2O_4$, considering slabs of 8.5 Å and 17 Å thickness. We found that, after the surface reconstruction, the Zn2 termination is the stable one. We also found that cationic inversion, a defect in volumetric $ZnFe_2O_4$, become energetically favourable in the surface. This result explains the high degree of inversion observed in nano-sized and thin films of $ZnFe_2O_4$. The presence of some inversion degree in the samples gives rise to $Fe^A$–O–$Fe^B$ coupling (being $Fe^A$ and $Fe^B$ Fe atoms located at sites A and B of the spinel structure) that strengthen the magnetic interactions and the ferrimagnetic behavior at high temperatures. Calculations of the magnetic coupling constants in normal and inverted bulk and thin films of $ZnFe_2O_4$ are now in progress in order to correlate the degree of inversion with the magnetic behavior of this semiconducting oxide.

**Acknowledgments.**

This research was partially supported by CONICET (Grant No. PIP0747, PIP0720), UNLP (Grant No. 11/X678, 11/X680, 11/X708, 11/X788, 11/X792), ANPCyT (Grant No. PICT PICT 2012-1724, 2013-2616, 2016-4083) and UNNOBA (Grant No. SIB0176-2017), and "Proyecto Acelerado de Cálculo 2017", Red Nacional de Computación de Alto Desempeño (SNCAD-MINCyT) - HPC Cluster, Rosario. Argentina. Funded by the Deutsche Forschungsgemeinschaft (DFG, German Research Foundation) –Projektnummer 31047525- SFB 762, Projects A4 and B1.

|  |  | $N_{Zn}$ | $N_{Fe}$ | $N_O$ |
|---|---|---|---|---|
| 8.5 Å thickness slab | Zn2 termination | 10 | 16 | 32 |
|  | O4-Fe4 termination | 8 | 20 | 32 |
|  | O4-Fe4-O4 termination | 8 | 20 | 40 |
| 17 Å thickness slab | Zn2 termination | 18 | 32 | 64 |
|  | O4-Fe4 termination | 16 | 36 | 64 |
|  | O4-Fe4-O4 termination | 16 | 36 | 72 |
|  |  |  |  |  |

Table 1: Number of Zn, Fe, and O atoms ($N_{Zn}$, $N_{Fe}$, and $N_O$) in the slabs used for the simulation of the different (001) surface terminations of $ZnFe_2O_4$.

|  |  | $\gamma$ (meV/A$^2$) O4-Fe4-O4 termination | $\gamma$ (meV/A$^2$) O4-Fe4 termination | $\gamma$ (meV/A$^2$) Zn2 termination |
|---|---|---|---|---|
| O-rich condition | | | | |
| FP-LAPW, 8.5 Å thickness slab | Normal | 310 | 490 | -120 |
| | Superficial antisite | 310 | 490 | -120 |
| | Deep antisite | 310 | 500 | -100 |
| QE, 8.5 Å thickness slab | Normal | 218 | 386 | 155 |
| | Superficial antisite | 218 | 386 | 158 |
| | Deep antisite | 216 | 395 | 202 |
| QE, 17 Å thickness slab | Normal | 222 | 390 | 158 |
| | Superficial antisite | 223 | 390 | 161 |
| | Deep antisite | 222 | 400 | 204 |
| O-poor condition | | | | |
| FP-LAPW, 8.5 Å thickness slab | Normal | 380 | 260 | -190 |
| | Superficial antisite | 380 | 260 | -190 |
| | Deep antisite | 380 | 270 | -170 |
| QE, 8.5 Å thickness slab | Normal | 293 | 159 | 80 |
| | Superficial antisite | 293 | 159 | 83 |
| | Deep antisite | 292 | 167 | 127 |
| QE, 17 Å thickness slab | Normal | 298 | 163 | 83 |
| | Superficial antisite | 297 | 163 | 86 |
| | Deep antisite | 298 | 172 | 128 |

Table 2: Predicted surface energy formations $\gamma$ (in meV/A$^2$) for the different terminations and slabs studied here by means of FP-LAPW and QE calculations.

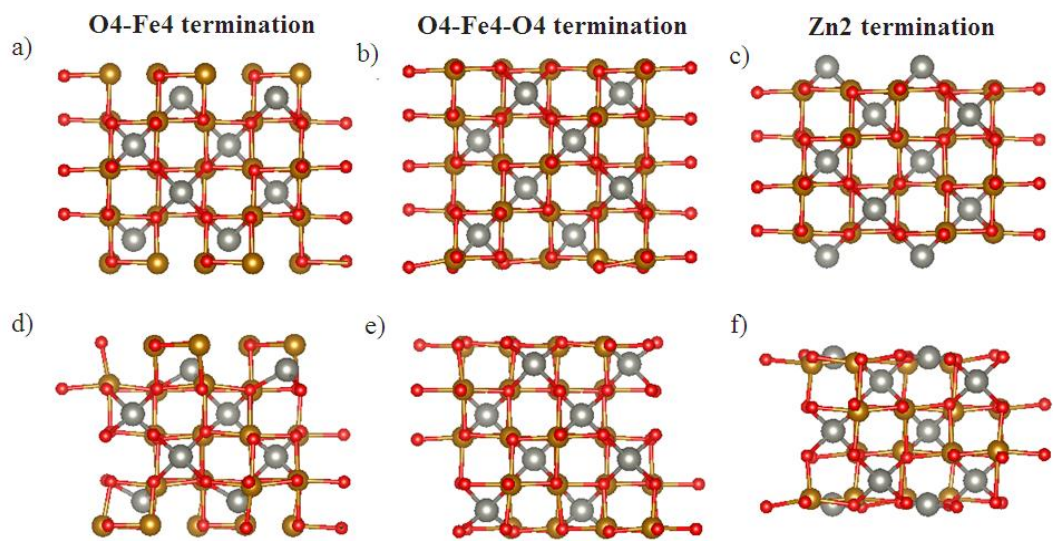

Figure 1: Slab models for the different (001) surface terminations of ZnFe$_2$O$_4$. Top: Initial (un-reconstructed) structures of ZFO (001) surface terminations. Bottom: Reconstructed structures.

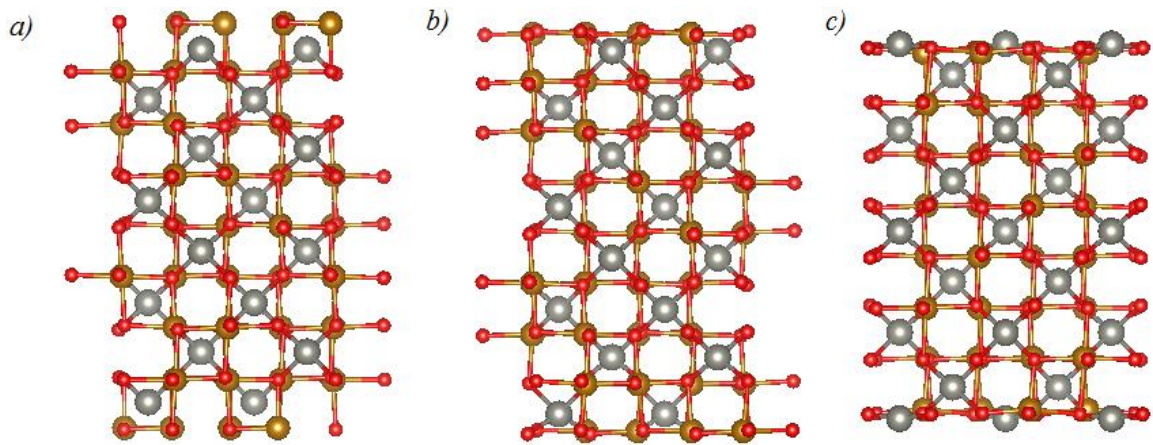

Figure 2: 17 Å thickness slab model of $ZnFe_2O_4$ (001) surface terminations. a) O4-Fe4 termination. b) O4-Fe4-O4 termination. c) Zn2 termination. Only the reconstructed terminations are shown.

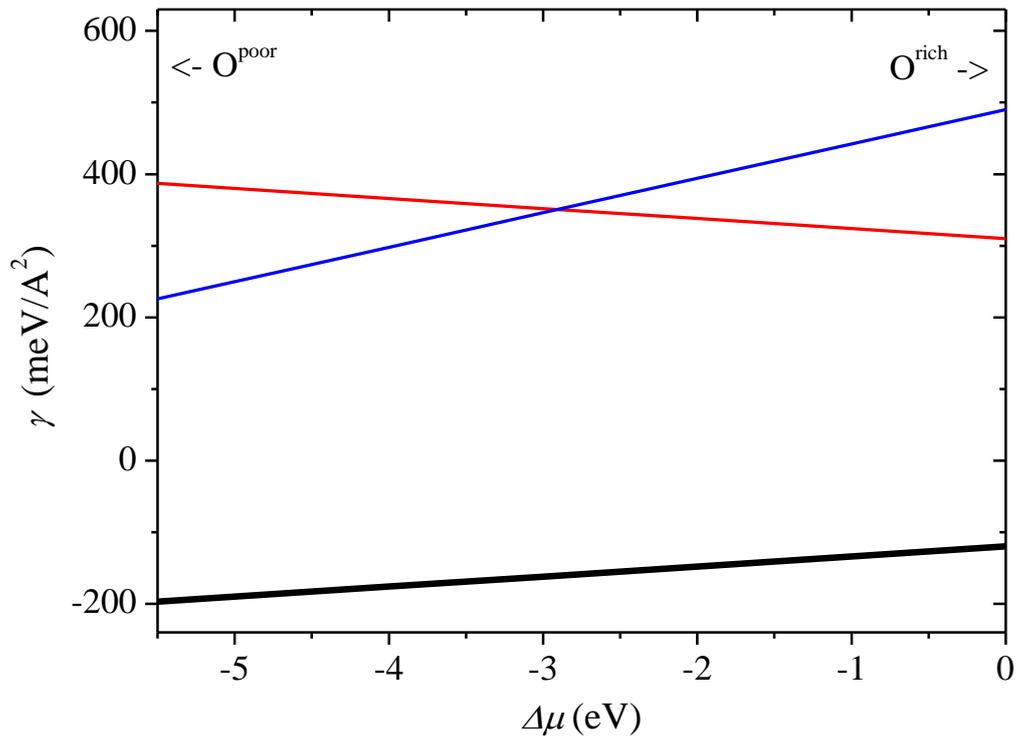

Figure 3: Calculated surface energies ($\gamma$) for the three (001) terminations of $ZnFe_2O_4$ studied here. Red, blue, and black lines represent the surface energies of the terminations O4-Fe4-O4, O4-Fe4 and Zn2 terminations, respectively. The left and right x-axis ends correspond to the O-poor and O-rich environments. The calculations correspond to the FP-LAPW calculations, 8,5 Å thickness slab.

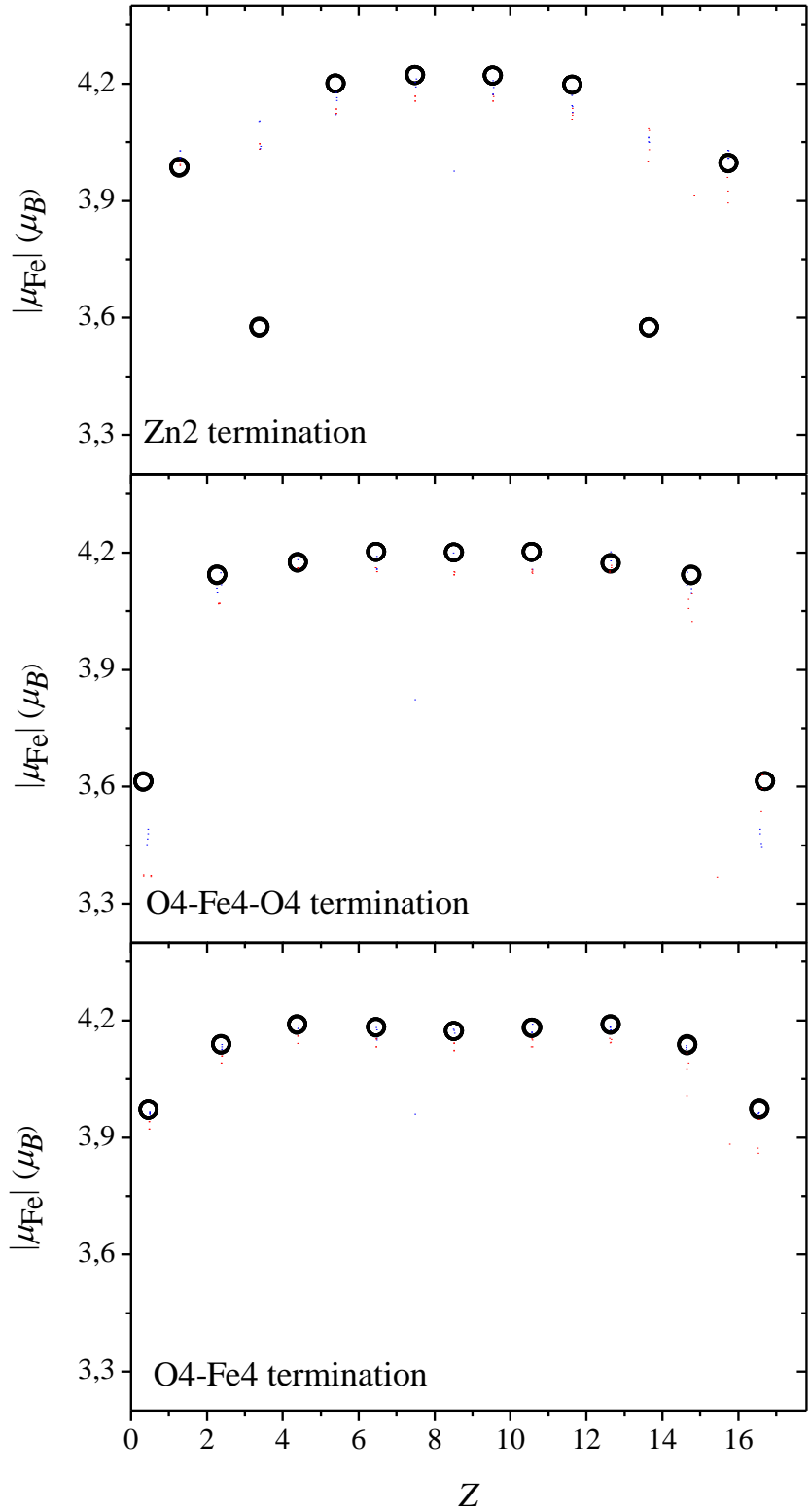

Figure 4: Magnetic moments $\mu$ at the Fe sites as a function of the distance (in Å) to the surface ($Z$) for the three (001) surface terminations studied here. In all cases the results correspond to the normal structures. $Z=0$ correspond to the lower layer of the slabs showed in Figure 1.